\documentstyle[twoside,epsf]{article}

\catcode`\@=11
\long\def\@makefntext#1{
\protect\noindent \hbox to 3.2pt {\hskip-.9pt
$^{{\eightrm\@thefnmark}}$\hfil}#1\hfill}       

\def\@makefnmark{\hbox to 0pt{$^{\@thefnmark}$\hss}}    

\def\ps@myheadings{\let\@mkboth\@gobbletwo
\def\@oddhead{\hbox{}
\rightmark\hfil\eightrm\thepage}
\def\@oddfoot{}\def\@evenhead{\eightrm\thepage\hfil
\leftmark\hbox{}}\def\@evenfoot{}
\def\sectionmark##1{}\def\subsectionmark##1{}}

\newcounter{sectionc}\newcounter{subsectionc}\newcounter{subsubsectionc}
\renewcommand{\section}[1] {\vspace{12pt}\addtocounter{sectionc}{1}
\setcounter{subsectionc}{0}\setcounter{subsubsectionc}{0}\noindent
    \par\vspace{5pt}}
\renewcommand{\subsection}[1] {\vspace{12pt}\addtocounter{subsectionc}{1}
    \setcounter{subsubsectionc}{0}\noindent
    {\bf\thesectionc.\thesubsectionc. {\kern1pt \bfit #1}}\par\vspace{5pt}}
\renewcommand{\subsubsection}[1] {\vspace{12pt}\addtocounter{subsubsectionc}{1}
    \noindent{\tenrm\thesectionc.\thesubsectionc.\thesubsubsectionc.
    {\kern1pt \tenit #1}}\par\vspace{5pt}}

\newcounter{appendixc}
\newcounter{subappendixc}[appendixc]
\newcounter{subsubappendixc}[subappendixc]
\renewcommand{\thesubappendixc}{\Alph{appendixc}.\arabic{subappendixc}}
\renewcommand{\thesubsubappendixc}
        {\Alph{appendixc}.\arabic{subappendixc}.\arabic{subsubappendixc}}

\renewcommand{\appendix}[1] {\vspace{12pt}
        \refstepcounter{appendixc}
        \setcounter{figure}{0}
        \setcounter{table}{0}
        \setcounter{lemma}{0}
        \setcounter{theorem}{0}
        \setcounter{corollary}{0}
        \setcounter{definition}{0}
        \setcounter{equation}{0}
        \renewcommand{\thefigure}{\Alph{appendixc}.\arabic{figure}}
        \renewcommand{\thetable}{\Alph{appendixc}.\arabic{table}}
        \renewcommand{\theappendixc}{\Alph{appendixc}}
        \renewcommand{\thelemma}{\Alph{appendixc}.\arabic{lemma}}
        \renewcommand{\thetheorem}{\Alph{appendixc}.\arabic{theorem}}
        \renewcommand{\thedefinition}{\Alph{appendixc}.\arabic{definition}}
        \renewcommand{\thecorollary}{\Alph{appendixc}.\arabic{corollary}}
        \renewcommand{\theequation}{\Alph{appendixc}.\arabic{equation}}
   \noindent{\tenbf Appendix \theappendixc. #1}\par\vspace{5pt}}
\newcommand{\subappendix}[1] {\vspace{12pt}
        \refstepcounter{subappendixc}
        \noindent{\bf Appendix \thesubappendixc. {\kern1pt \bfit #1}}
        \par\vspace{5pt}}
\newcommand{\subsubappendix}[1] {\vspace{12pt}
        \refstepcounter{subsubappendixc}
        \noindent{\rm Appendix \thesubsubappendixc. {\kern1pt \tenit #1}}
        \par\vspace{5pt}}

\newcommand{\textlineskip}{\baselineskip=13pt}
\newcommand{\smalllineskip}{\baselineskip=10pt}

\def\eightcirc{
\begin{picture}(0,0)
\put(4.4,1.8){\circle{6.5}}
\end{picture}}
\def\eightcopyright{\eightcirc\kern2.7pt\hbox{\eightrm c}}

\def\abstracts#1#2#3{{
    \centering{\begin{minipage}{4.5in}\baselineskip=10pt\footnotesize
    \parindent=0pt #1\par
    \parindent=15pt #2\par
    \parindent=15pt #3
    \end{minipage}}\par}}

\renewenvironment{thebibliography}[1]
    {\frenchspacing
     \ninerm\baselineskip=11pt
     \begin{list}{\arabic{enumi}.}
        {\usecounter{enumi}\setlength{\parsep}{0pt}
     \setlength{\leftmargin 12.7pt}{\rightmargin 0pt} 
    \setlength{\leftmargin 17pt}{\rightmargin 0pt}   
    \setlength{\leftmargin 22pt}{\rightmargin 0pt}   
         \setlength{\itemsep}{0pt} \settowidth
    {\labelwidth}{#1.}\sloppy}}{\end{list}}

\newcounter{itemlistc}
\newcounter{romanlistc}
\newcounter{alphlistc}
\newcounter{arabiclistc}

\newcommand{\fcaption}[1]{
        \refstepcounter{figure}
        \setbox\@tempboxa = \hbox{\footnotesize Fig.~\thefigure. #1}
        \ifdim \wd\@tempboxa > 5in
           {\begin{center}
        \parbox{5in}{\footnotesize\smalllineskip Fig.~\thefigure. #1}
            \end{center}}
        \else
             {\begin{center}
             {\footnotesize Fig.~\thefigure. #1}
              \end{center}}
        \fi}

\def\@citex[#1]#2{\if@filesw\immediate\write\@auxout
    {\string\citation{#2}}\fi
\def\@citea{}\@cite{\@for\@citeb:=#2\do
    {\@citea\def\@citea{,}\@ifundefined
    {b@\@citeb}{{\bf ?}\@warning
    {Citation `\@citeb' on page \thepage \space undefined}}
    {\csname b@\@citeb\endcsname}}}{#1}}

\newif\if@cghi
\def\citelow{\@cghifalse\@ifnextchar [{\@tempswatrue
    \@citex}{\@tempswafalse\@citex[]}}

\def\@refcitex[#1]#2{\if@filesw\immediate\write\@auxout
    {\string\citation{#2}}\fi
\def\@citea{}\@refcite{\@for\@citeb:=#2\do
    {\@citea\def\@citea{, }\@ifundefined
    {b@\@citeb}{{\bf ?}\@warning
    {Citation `\@citeb' on page \thepage \space undefined}}
    \hbox{\csname b@\@citeb\endcsname}}}{#1}}

\def\@refcite#1#2{{#1\if@tempswa\typeout
        {IJCGA warning: optional citation argument
    ignored: `#2'} \fi}}

\def\refcite{\@ifnextchar[{\@tempswatrue
    \@refcitex}{\@tempswafalse\@refcitex[]}}

\def\pmb#1{\setbox0=\hbox{#1}
    \kern-.025em\copy0\kern-\wd0
    \kern.05em\copy0\kern-\wd0
    \kern-.025em\raise.0433em\box0}

\def\fnt#1#2{\footnotetext{\kern-.3em
    {$^{\mbox{\scriptsize #1}}$}{#2}}}

\def\runninghead#1#2{\pagestyle{myheadings}
\markboth{{\protect\footnotesize\it{\quad #1}}\hfill}
{\hfill{\protect\footnotesize\it{#2\quad}}}}
\font\tenrm=cmr10
\font\tenit=cmti10
\font\tenbf=cmbx10
\font\bfit=cmbxti10 at 10pt
\font\ninerm=cmr9

\font\eightrm=cmr8

\def\qed{\hbox{${\vcenter{\vbox{            
   \hrule height 0.4pt\hbox{\vrule width 0.4pt height 6pt
   \kern5pt\vrule width 0.4pt}\hrule height 0.4pt}}}$}}

\begin{document}

\newpage

\runninghead{\tiny Rosu, Planat, Saniga} {\tiny MUBs: From FPG to QPE}

\normalsize\textlineskip
\thispagestyle{empty}
\setcounter{page}{1}


\vspace*{0.88truein}

\bigskip
\centerline{\large \bf MUBs: From Finite Projective Geometry to Quantum Phase Enciphering}

\vspace*{0.035truein}
\vspace*{0.37truein}
\vspace*{10pt} \centerline{\footnotesize H. C.
ROSU\footnote{\tiny E-mail: hcr@ipicyt.edu.mx $\quad$ mubg04.tex \hfill Status: accepted at AIP Conf. QCMC 2004, Strathclyde, Glasgow}, M. PLANAT, M. SANIGA }
\vspace*{0.015truein}
\centerline{\footnotesize  Applied Math-IPICyT, Institut FEMTO-ST, Slovak Astronomical Institute}
\centerline{\tiny [Mexico-Romania, France, Slovak Republic]}
\centerline{\small quant-ph/0409096}
\vspace*{0.225truein}


\vspace*{0.21truein} \abstracts{ {\bf Abstract.} This short note highlights the most prominent mathematical problems and physical questions associated with the 
existence of the maximum sets of mutually unbiased bases (MUBs) in the Hilbert space of a given dimension. }{}{}


\textlineskip                  
\vspace*{12pt}                 

\vspace*{1pt}\textlineskip  
\vspace*{-0.5pt}
\noindent


\noindent






{\bf Introduction}

\medskip

\noindent
Technical problems in quantum information theory already connect
such distinct disciplines as number theory, abstract algebra and
projective geometry. For a partial list of open problems related
to the development of quantum computing technologies, see
{\em http://www.imaph.tu-bs.de/qi/problems}. In this stimulating
research area the issue of the so-called mutually unbiased bases
is an important one since it is related  to the complete state
determination of a quantum system.

\medskip

{\bf Definitions and Basic Facts}

\medskip

\noindent
Two different orthonormal bases $A$ and $B$ of a $d$-dimensional
Hilbert space are called {\em mutually unbiased} if and only if
\begin{equation}
|\langle a|b\rangle|=1/\sqrt{d}~,
\end{equation}
for all $a \in A$ and all $b\in B$. An aggregate of MUBs is a set
of orthonormal bases which are pairwise mutually unbiased. It has
been found that the maximum number of such bases cannot be greater
than $d+1$ in $d$-dimensional Hilbert space \cite{1.}. It is also
known that this limit is reached if $d$ is a power of a prime.

Yet, a still unanswered question is if there are non prime power
values of $d$ for which this bound is attained. Based on numerical
calculations, it is generally agreed \cite{2.} that in the
latter cases the lower bound on the maximum number of such bases is
$$
 N_{\rm max}= 1+{\rm min} (p_i^{e_i}),
$$
where ${\rm min} (p_i^{e_i})$ is the lowest factor in the prime
number decomposition of $d=\prod _i p_{i}^{e_i}$.

Whether or not there exists a set of $d+1$ MUBs in a
$d$-dimensional Hilbert space if $d$ not a power of a prime could
be intimately linked with the question of the existence of
projective planes whose order is not a power of prime according to
a conjecture that we published recently \cite{3.}. Another interesting
recent result is that when the quantum measurements are performed in
MUBs, a simple linear and universal relation exists between the
post-measurement density matrix and the pre-measurement density matrix
\cite{4.}.

The main application of MUBs pertains to secure quantum key
exchange (quantum cryptography). This is because any attempt by an
eavesdropper (say Eve) to distinguish between two non-orthogonal
quantum states shared by two remote parties (say Alice and Bob)
will occur at the price of introducing a disturbance to the
signal, thus revealing the attack, and allowing to reject the
corrupted quantum data. Modern protocols, e.g., the original BB84
protocol, use only one qubit technologies implying dimension
$d=2$, usually the polarisation states of the photon. But the
security against eavesdropping increases when all the three bases
of qubits are used, or by using qudits, or entanglement-based
protocols.

Quantum state recovery and secure quantum key distribution can
also be achieved using positive operator valued measures (POVMs)
which are symmetric informationally complete (SIC-POVMs)
\cite{5.}. These are sets of $d^2$ normalized vectors $a$ and $b$ such
that
$$
 |\langle a|b\rangle|=  1/(d + 1)^{1/2}\,\, {\rm   when} \,\, a \neq b~.
$$
Unlike the MUBs, the SIC-POVMs could exist in all finite
dimensions. Recently, SIC-POVMs have been constructed  in
dimension $d=6$ \cite{6.}.

\medskip

{\bf MUB's and Finite Projective Planes}

\medskip

\noindent
As already mentioned, we have recently conjectured \cite{3.} that the
existence of the maximum set of MUBs in a given dimension $d$ and that
of a projective plane of the same dimension may well represent two
aspects of one and the same problem. Perhaps the most serious backing
of our surmise is found in a recent paper by Wootters
\cite{7.}. Associating a line in a finite geometry with a pure state
in the quantum problem, the author shows that a complete set of MUBs
is, in some respects, analogous to a finite {\it affine} plane, and
another kind of quantum measurement, the SIC-POVM, is also analogous
to the same configuration, but with the swapped roles of points and
lines. It represents no difficulty to show that this ``dual" view of
quantum measurement is deeply rooted in our conjecture. To this end,
it suffices to recall two facts \cite{8.}. First, any affine plane is
a particular subplane (subgeometry) of a projective plane, viz. a
plane which arises from the latter if one {\it line}, the so-called
``line at infinity," is deleted. Second, in a projective plane, there
is a perfect {\it duality} between points and lines; that means, to
every projective plane, $S_{2}$, there exists a dual projective plane,
$\Sigma_{2}$, whose points are the lines of $S_{2}$ and whose lines
are the points of $S_{2}$ \cite{9.}. So, {\it affinizing} $S_{2}$
means deleting a {\it point} of $\Sigma_{2}$ and thus, in light of our
conjecture, qualitatively recovering the results of Wootters, shedding
also important light on some other of the most recent findings
\cite{4.,10.}. The latter reference, in fact, gives several strong
arguments that there are no more than three MUBs in dimension six, the
latter being the smallest non-prime-power dimension.

\medskip

{\bf MUB's and Quantum Fourier Transforms (QFT's)}

\medskip

\noindent
There is a useful relationship between MUBs and QFTs.  Consider a
basis B$_0$ =$(|0\rangle, |1\rangle,...,|d\,-1\rangle )$ with
indices $n$ in the ring $Z_{d}$ of integers modulo $d$. The dual
basis defined by the quantum Fourier transform is
$$
|\theta _k\rangle =\frac{1}{\sqrt{d}}\sum _{n=0}^{d\,-1} \omega
_{d} ^{nk}|n\rangle~, \quad \qquad \omega _{d}
=\exp\left(i\frac{2\pi}{d}\right).
$$
In the particular case $d=2$, $\omega _{\rm 2} =-1$ and the
$\theta$ basis acquires the form
$$
|\theta _0\rangle =\frac{1}{\sqrt{2}}(|0\rangle + |1\rangle)~;
\quad  |\theta _1\rangle = \frac{1}{\sqrt{2}}(|0\rangle -
|1\rangle).
$$
Note that the two orthogonal bases B$_0$ =$(|0\rangle, |1\rangle)$
and B$_1$ =$(|\theta _0\rangle, |\theta _1\rangle)$ are mutually
unbiased. The third base B$_2$ =$(|\psi _0\rangle, |\psi
_1\rangle)$, mutually unbiased to them, is obtained from $H$ (the
Hadamard matrix) amended by the action to the right of a $\pi$/2
rotation $S$
$$
H=\frac{1}{\sqrt{2}}\left( \begin{array}{cc}
1 & 1 \\
1 & -1\end{array} \right )~, \quad
S=\left( \begin{array}{cc}
1 & 0 \\
0 & i\end{array} \right ) \qquad  \rightarrow \qquad HS= \frac{1}{\sqrt{2}}\left( \begin{array}{cc}
1 & i \\
1 & -i\end{array} \right )
$$
as applied to B$_0$.

The fact that the B bases for $d=2$ are also the eigenvectors of
the Pauli matrices $\sigma _z$, $\sigma _x$, and $\sigma _y$,
respectively, has led to a method of MUBs' construction in
dimension $d$ in terms of the generalized Pauli operators
$$
X_{d}|n\rangle = |n+1\rangle~, \qquad Z_{d}|n\rangle = \omega
_{d}^{n} |n\rangle,
$$
known as shift and clock operators. For a prime dimension $d=p$,
it can be shown \cite{11.} that the eigenvectors of the set of unitary
operators $Z_{p}, X_{p}, X_{p}Z_{p}, ..., X_{p}Z_{p}^{p-1}$ generate
the corresponding $d+1$ MUBs.

Can MUBs be obtained for any $d$ and any (number) field by Fourier
transforms as in the case of $d=2$?  In principle, the answer is
yes. For this, one should employ such a quantum Fourier transform
whose exponent $\omega$ now acts on a finite (Galois) field,
$G=GF(p^m)$, having characteristic $p$ and $d=p^m$ elements.
Denote ``$\oplus$"  and ``$\bullet$" the two operations in the
field, corresponding to ``+" and ``$\cdot$" in the field of real
numbers. Then, the $G$-Fourier transform reads
$$
|\theta _k\rangle =\frac{1}{\sqrt{d}}\sum _{n=0}^{d\,-1} \omega
_{p}^{n\bullet k}|n\rangle~.
$$
Given any two polynomials $k$ and $n$ in $G$, there exists a
uniquely determined pair $a$ and $b$ in $G$ such that
$$
k=a \bullet n \oplus b,
$$
where deg $a$ $>$ deg $b$, so that the exponent in the $G$-quantum
Fourier transform can be written in  the form
$$
E=n\bullet (a\bullet n\oplus b).
$$
The last formula is valid for the case of a prime dimension $d=p$
for which $E$ is an integer. Otherwise, it has to be replaced by
the trace of $GF(p^m$), i.e., a map down to $GF(p)$ defined as
follows
$$
tr(E) = E +E^{p}+... +E^{p^{m-1}}~, \quad E\in GF(p^m),
$$
and so
$$
|\theta _b^a\rangle =\frac{1}{\sqrt{d}}\sum _{n=0}^{d \,-1} \omega
_{p} ^{tr [n \bullet (a \bullet n \oplus b)]}|n\rangle~.
$$
This general formula was first obtained by Wootters \& Fields and
further insights into it have recently been given \cite{12.}. In a
Galois field of odd characteristic, the latter formula provides us
with a set of $d$ bases of index $a$ for the base and index $b$ for
the vector in the base, mutually unbiased to each other as well as to
the computational base B$_0$. It is worth noting that this strategy of
constructing MUBs fails for characteristic two, since in this case
$$
|\sum _{n=0}^{d \,-1} \omega _{2} ^{tr [n \bullet (a \bullet n
\oplus b)]}|n \rangle|=0~,
$$
irrespectively of the values of $a$ and $b$. In this case, one has
to use Galois rings instead of Galois fields in order to find MUBs
\cite{12.,13.}.

Finally, we notice that the same formula provides an interesting
relationship between MUBs and quantum phase operators
\cite{14.}. Indeed, it is known that the Fourier basis $|\theta
_k\rangle$ can be derived as the set of eigenvectors of a quantum
phase operators $\Theta =\sum _{k=0}^{d-1} \theta _k|\theta _k\rangle
\langle \theta _k|$. Thus, viceversa, each base of index $a$ can be
associated to a quantum phase operator of the form
$$
\Theta ^a= \sum _{b=0}^{d-1}\theta _b^a|\theta _b^a\rangle \langle \theta _b ^a|~.
$$
The implementation of the MUB concept at the level of quantum
phase kets and operators could have important technological
consequences for defining generalized measurements of the quantum
phase, which is a key feature in quantum computing processes.

\medskip

{\bf Conclusion}

\medskip

\noindent
This short note highlights only the most prominent mathematical
problems and physical questions associated with the existence of
the maximum sets of MUBs in the Hilbert space of a given
dimension. Yet, it should give the reader a fairly good picture of
the state of the art of the topic and why the latter entails
steadily-increasing attention of both physicists and
mathematicians.


\end{document}